# Depth-first search in split-by-edges trees

Asbjørn Brændeland

***Abstract***

A *split-by-edges* tree of a graph $G$ is a binary tree $T$ where the root = $V(G)$, every leaf is an independent set in $G$, and for every other node $N$ in $T$ with children $L$ and $R$ there is an edge $uv$ in $G$ such that $u$ and $v$ are in $N$, $L = N - v$ and $R = N - u$. Every maximum independent set in $G$ is in the same layer $M$, which is guarantied to be found in a layer by layer search of $T$. Every depth-first search in $T$ will reach $M$, and may stop there, if a leaf was hit, or will pass through, to another, smaller, leaf below. For random graphs of a given order and increasing size the succes rates of depth-first searches in split-by-edges trees oscillate between local minima and maxima.

The *split by edges tree* and the *uniquified split by edges tree* are defined in [1] as follows.

**Definition 1**: Let $G$ be graph and let $T$ be a binary tree of subsets of $V(G)$. Then $T$ is a **split-by-edges tree**, or **SBE-tree**, of $G$ if and only if the root of $T = V(G)$, every leaf in $T$ is an independent set of $G$, and for every other node $N$ in $T$ with children $L$ and $R$ there is a pair of vertices $\{u, v\} \subset N$ such that $L = N - u$, $R = N - v$, and $u$ and $v$ are adjacent in $G$.

**Definition 2**: An SBE-tree minus its duplicate nodes is a **uniquified SBE-tree**, or a **USBE-tree**.

By Corollary 1.1. in [1], which says that *every maximal independent set of G is a leaf in every SBE-tre of G*, there is a layer $M_I$ in $T$ that contains every maximum independent set in $G$[1], and the distance from the root of $T$ to $M_I$ is $n - \alpha(G)$, when $n = |V(G)|$, and $\alpha(G)$ is the independence number of $G$; and it follows from definition 2 that there is a corresponding layer $M'_I$ in the uniquified version $T'$ of $T$. For random graphs a *layer-by-layer search*, which is akin to a *breadth-first search*, through $T'$ produces $M'_I$ in maximum $2^{O(0.37n)}$ time [1]. A *depth-first search* through $T$ reaches $M_I$ in $n - \alpha(G)$ time, but since $M_I$ usually also contains non-independent sets, it is far from guarantied that an independent set will be hit, and if not, the search continues until a smaller independent set is found further down.

Whether the SBE-tree being searched is uniquified or not, depends on the search method. In a layer-by-layer search the layers are created by the search procedure, which can keep duplicates out of the way. But since the cardinalities of the nodes are unique to each layer, duplicates can only occur layerwise, thus in a depth-first search, where only one node or two sibling nodes are examined per layer, the existence of duplicates is not relevant—which is to say that the tree being searched is not uniquified.

The vertices of a graph $G$ can be ordered by degree[1]. A layer-by-layer search for a maximum independent set in a USBE-tree of $G$ is more efficient if the order is descending and less efficient if the order is ascending than if the order is arbitrary, due to the fact that a descending order gives a slimmer USBE-tree. Somewhat surprisingly, the effect is the opposite for depth-first search. The real surprise is that there is an effect at all, since in an SBE-tree the width of layer $L_l = 2^l$ all the way down to the layer that contains the maximum independent sets, regardless of how the graph is arranged.

Independent of vertex ordering, the DFS success rate increases if we, instead of always choosing the left branch, choose the one with the *fewest edges*, or even better, the *most stable* branch, when the stability of a node $N$ on $n$ vertices is given by the formula

$$\sum_{v \in N} \frac{n}{\deg_{G(N)}(v) + 1} \quad (1)$$

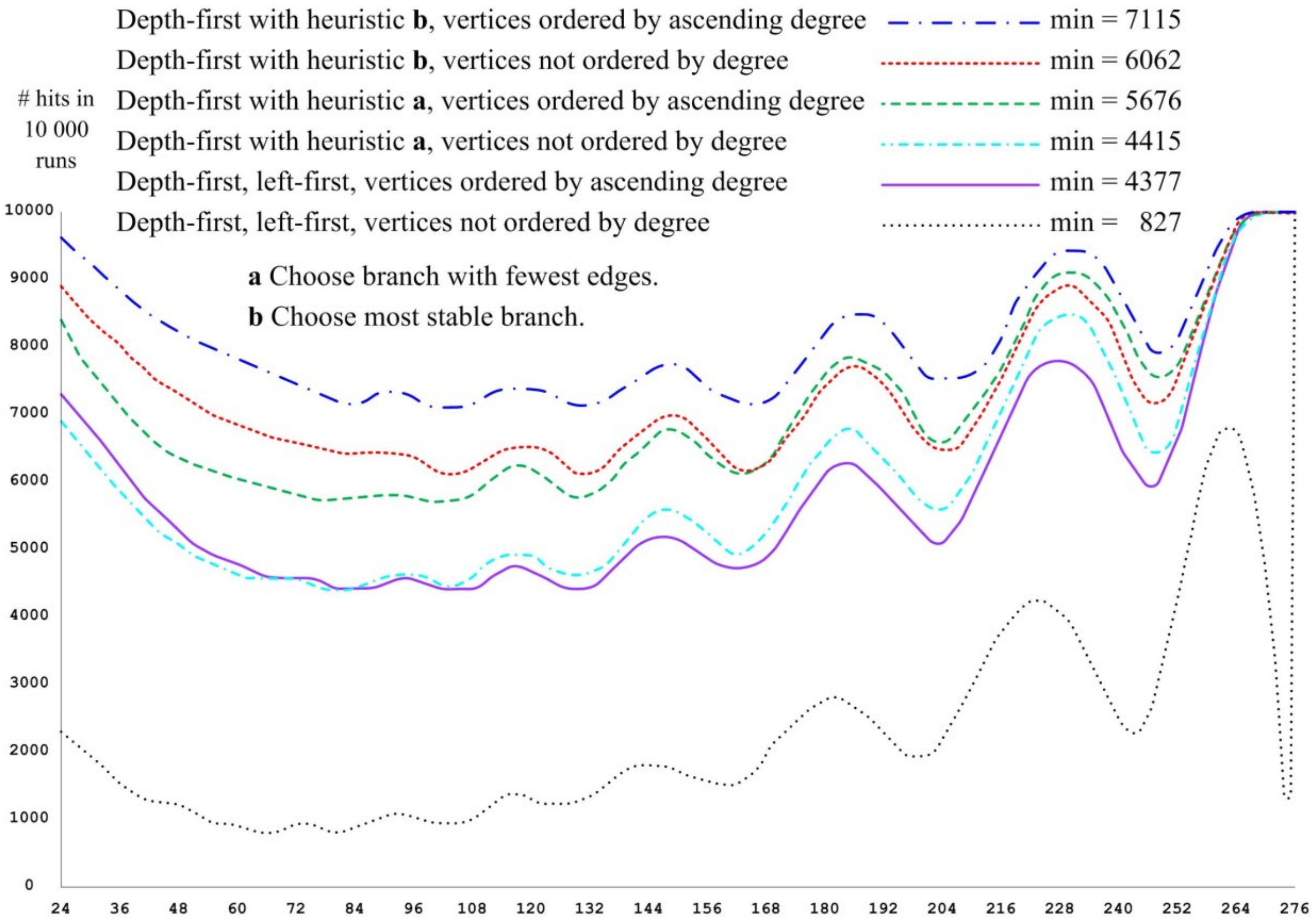

Figure 1. The curves show the number of successful MIS searches for six variations of depth-first search in the SBE trees of random graphs on 24 vertices. 10 000 graphs where searched for each size ranging from 24 to 276. (That there for some *m* are less than 10 000 different graphs, has no bearing on the shapes of the curves.)

We will return to the peculiar oscillations shown in Figure 1, presently.

The curves keep their shapes as *n* grows, but the success rates drop rather quickly, as shown in Figure 2.

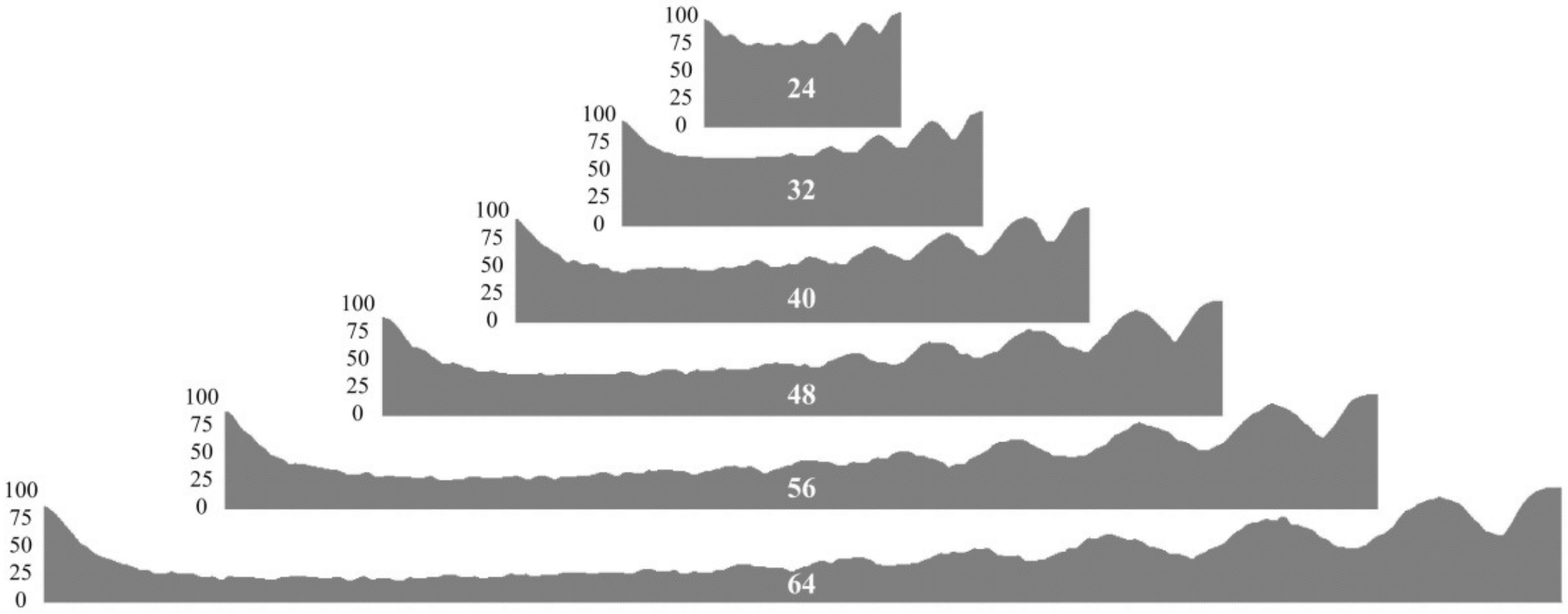

Figure 2. The histograms show the number of successful *depth-first*, *most-stable-branch* MIS searches in the SBE trees of six random graphs of order 24 to 64. 100 graphs where searched for each size, ranging from $n$ to $n(n-1)/2$.The minimum success ratio drops from about 0.71, for $n = 24$, to about 0.25, for $n = 64$.

The *accuracy* of an MIS algorithm $A$ for a graph $G$ is the ratio $A(G) / \alpha(G)$, when $A(G)$ is the result of applying $A$ to $G$. Accuracy is subject to the same size related oscillations as success. For $n = 24$ to 64 the lowest accuracy of depth-first, most-stable-branch searches in SBE-trees falls from 0.917 to 0.841. (A different, but nearly equivalent, type of measure is defined in [2].)

In Figure 1 and Figure 2, the numbers of successful depth-first searches for random graphs of a fixed order and increasing size oscillate between local minima and maxima. This oscillation shows up consistently in every comparable test, including tests where different random graph generators were used. In the grids on top of Figure 3 the red lines and the blue dottet lines indicate local maxima and minima, respectively, for depth-first searches in graphs of orders 15 to 40. The grid points can be seen as coordinates $(n, m)$ of the graph orders and sizes that give the local extrema. The grids are based on 100 runs for each $(n, m)$[1], and the vertical lines have been slightly smoothed.

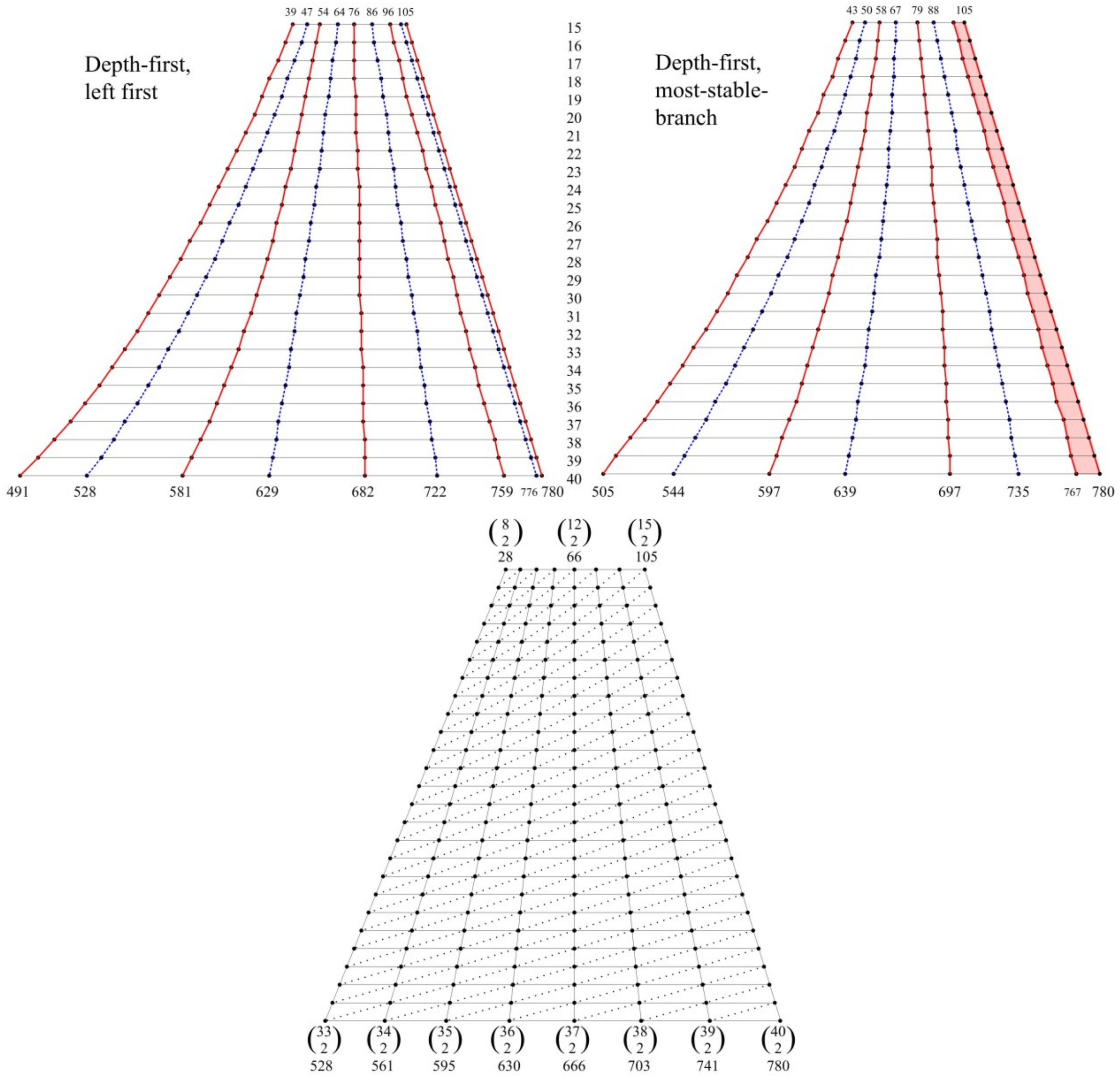

Figure 3.

For comparison Figure 3 also contains a grid of $\binom{n}{2}$ values, where each dotted line indicates *one* value. Each vertical line contains a segment of the partial sums of $N = (1, \ldots, 40)$. Conversely, the difference sequences of each vertical line is a segment of $N$. The corresponding difference sequences for the two upper grids are close to some, but not quite equal to any segments of $N$.

[1] The fact that there is only one graph with $n$ vertices and $n(n-1)/2$ edges is immaterial, since no algorithm ever fails for this value of $m$.